\documentstyle[a4,12pt]{article}

\newcommand{\NP}[1]{Nucl. Phys.\ {\bf #1}\ }
\newcommand{\PL}[1]{Phys. Lett.\ {\bf #1}\ }

\newcommand{\PRL}[1]{Phys. Rev. Lett.\ {\bf #1}\ }

\newcommand{\JMP}[1]{J. Mod. Phys.\ {\bf #1}\ }
\newcommand{\MPL}[1]{Mod. Phys. Lett.\ {\bf #1}\ }
\newcommand{\CQG}[1]{Class. Quantum Grav.\ {\bf #1}\ }

\def\sqr#1#2{{\vcenter{\hrule height.#2pt
     \hbox{\vrule width.#2pt height#1pt \kern#1pt
        \vrule width.#2pt}
     \hrule height.#2pt}}}

\def\square{\mathchoice\sqr68\sqr68\sqr{4.2}6\sqr{3.0}6}

\def\thinspace{\kern .16667em}

\def\Dir{\nabla\kern-7.8pt\Big{/}}

\def\ww{\wedge}
\def\part{\partial}

\def\aa{\alpha}

\def\dd{\delta}
\def\ee{\epsilon}
\def\ff{\phi}
\def\gg{\gamma}

\def\kk{\kappa}
\def\ll{\lambda}

\def\oo{\omega}
\def\OO{\Omega}

\def\pp{\psi}

\def\rr{\rho}

\def\ss{\sigma}
\def\SS{\Sigma}

\def\zz{\zeta}

\begin{document}
\begin{titlepage}
\begin{flushright}
CERN-TH 6286/91\\
DFTT 42/91\\
\end{flushright}
\vspace*{0.5cm}
\begin{center}
{\bf
\begin{Large}
{\bf HETEROTIC INSTANTONS AND SOLITONS \\ IN ANOMALY-FREE SUPERGRAVITY\\}
\end{Large}
}
\vspace*{2cm}
         {\large I. Pesando}\footnote{Work supported in part by
         Ministero dell'Universit\`{a} e della Ricerca Scientifica
         e Tecnologica}
         \footnote{email I\_PESANDO@TO.INFN.IT}
         \\[.5cm]
         Dipartimento di Fisica Teorica dell'Universit\`{a} di Torino\\
         Instituto Nazionale di Fisica Nucleare, Sezione di Torino\\
         via P.Giuria 1, I-10125 Turin, Italy \\[1cm]
         {\large A. K. Tollst\'{e}n}\footnote{email TOLLSTEN@CERNVM}
         \\[.5cm]
         CERN, CH-1211 Geneva 23, Switzerland
\end{center}
\vspace*{2cm}
\begin{abstract}
We extend the classical heterotic instanton solutions to all orders
in $\alpha'$ using the equations of anomaly-free supergravity, and
discuss the relation between these equations and the string theory
$\beta$-functions.
\end{abstract}
\vfill
CERN-TH 6286/91\\
DFTT 42/91\\
October 1991
\end{titlepage}
\section{Introduction}
During the past few years the study of classical (low energy) equations
for 10-dimensional superstrings
has yielded a number of interesting solutions, see for instance
\cite{DGHR90,S90,CHS91,C91,HaS91,HGS91,DL91}.
The standard method of finding these is to
solve the equations of motion
of ordinary supersymmetric Einstein-Yang-Mills theory
 (augmented by the Lorentz Chern-Simons term in the
definition of the antisymmetric tensor), which
equal the string $\beta$-functions to the lowest order.
To show that a solution obtained this way is also
a solution to string theory, one then tries to construct the
corresponding superconformal sigma-model.

However, from the space-time point of view it might appear somewhat
unsatisfactory to look for supersymmetric solutions using a set of
manifestly non-supersymmetric equations. A
supersymmetrization of the coupled Einstein-Yang-Mills
theory including the Lorentz Chern-Simons term
would also naturally extend the equations to higher orders in $\alpha'$,
which is interesting in itself. Such a supersymmetrization
has actually already been performed, both in the normal case (with a
3-form $H$)  \cite{BBLPT88,DaFRR88,P91}
 using an important observation by Bonora,
Pasti and Tonin \cite{BPT87}, and in the dual case \cite{N91}.
The equations of this model, which we call the Anomaly-Free
Supergravity (AFS), give $\alpha'$ corrections to the standard equations,
leading to implicit equations for the physical fields. In the general
case one must then expand to all orders in $\alpha'$.
AFS is hence a classically consistent, all orders in $\alpha'$ (on-shell)
supersymmetric theory which incorporates the Green-Schwarz condition for
anomaly cancellations \cite{GS84}.

Fortunately, for
the purpose of generalizing the classical solutions of
\cite{S90,CHS91}, it will turn
out to be sufficient to use the original implicit equations. This is what
is done in this paper. We find a solution to the AFS equations in a
closed form containing only $O(\alpha'^0)$ and $O(\alpha'^1)$ terms.
This solution, which is really a family of solutions containing the ones
by Callan, Harvey, and Strominger referred to above, consists of a
 non-linear differential equation
for the field appearing in the metric, which, in general,
might have to be solved as
an expansion in $\alpha'$. Also the relation between the dilaton and
the metric contains derivatives.

It should perhaps be pointed out here that although
AFS agrees with the effective string theory to lowest order and
contains terms to all orders,
it is extremely unlikely that it will turn out to
be equivalent to the massless effective string theory. It does not
incorporate the $\zeta(3)$ terms in an obvious fashion, and it has been
shown that AFS can at least in principle be extended to a non-minimal
version containing extra representations which can accommodate such
terms \cite{LP(T)}.
We believe, however, that the minimal AFS provides a better
approximation to string theory than the one normally used, and, as is
argued at the end of this paper, it might even provide a necessary
condition for a solution to be a solution of string theory.
\section{The instanton solution}
We will here study the generalization to AFS of the heterotic five-brane
solution by Strominger \cite{S90}.
We follow his calculation closely, only making a slightly more
general ansatz.
As in the lowest order case, the solution also turns out to incorporate
the wormhole solution \cite{CHS91,C91}
related to the solution with a five-brane source \`a la
Duff and Lu \cite{DL91}.
Since AFS corrections are rather complicated, we find it most
convenient to work directly in the variables of \cite{P91},
instead of
performing the field redefinitions \cite{S86,S90}
to obtain the $\ss$-model variables
and a flat five-brane metric.
We use mainly the conventions of \cite{DaFRR88,P91}.
For instance, $g_{MN}=(+,-,...,-)$, $\{\gg_M,\gg_N\}=2g_{MN}$ and a
p-form is defined as
$\oo^{(p)}=\oo_{M_1...M_p}~dz^{M_1}\ww...\ww dz^{M_2}$.
However, to make comparison easier, we use the index conventions of
\cite{S90},
that is, $M,N,P,...$ are 10-dimensional space-time indices,
 and $A,B,C...$ are
the corresponding tangent space indices.

We now want to find a maximally symmetric, supersymmetric solution, that
is, we want a solution to the
AFS equations with all spinorial fields equal
to zero, and with a (non-zero) Majorana-Weyl spinor $\ee$ satisfying
\cite{DaFRR88,P91}

$$\dd\pp_{M}=
  D_{M}\ee
  +{1\over36}\gg_{M}\gg_{A_1A_2A_3}\ee~T^{A_1A_2A_3}
  =0,
\eqno(1a)$$

$$\dd\ll=
  -2i~\gg^A\ee~\part_A\ff
  +i~\gg_{A_1A_2A_3}\ee~Z^{A_1A_2A_3}
  =0,
\eqno(1b)$$

$$\dd\chi=
  -{1\over4}\gg_{A_1A_2}\ee~F^{A_1A_2}
  =0.
\eqno(1c)$$

\noindent
To the lowest order, $T^{A_1A_2A_3}$ and $Z^{A_1A_2A_3}$ in the
gravitino and dilatino transformation equations are both proportional to
$H^{A_1A_2A_3}$.
This is no longer the case in AFS.
Here instead, we have the torsion
\footnote{Here, as in the following, we drop all fermionic terms.}

$$T_{A_1A_2A_3}=
  -3~e^{-{4\over3}\ff}H_{A_1A_2A_3}
  -2\gg_1~e^{-{4\over3}\ff}W_{A_1A_2A_3}
\eqno(2)$$

\noindent
where $\gg_1\sim\aa '$ and

$$W_{A_1A_2A_3}
  ={1\over2}\square\thinspace T_{A_1A_2A_3}
  +3~T_{B_1B_2[A_1}R^{B_1B_2}_{~~~~A_2A_3]}
  +3~T_{B_1[A_1A_2}R^{B_1}_{~~A_3]}
$$$$
  +4~T_{B_1B_2[A_1}T^{B_2~~B_3}_{~~A_2}T_{A_3]B_3}^{~~~~B_1}
  -({2\over27}+h_1)T_{A_1A_2A_3}T^2
\eqno(3)$$

\noindent
with $h_1$ a free parameter reabsorbable in a redefinition of
the dilaton:

$$e^{{4\over3}\ff^{'}}=
  e^{{4\over3}\ff}
  -2h_1\gg_1~T^2
\eqno(4)$$

\noindent
It is this torsion which turns up in covariant derivatives and the
curvature tensor;
$D=D(\OO)$, $R=R(\OO)$, and $\OO=\oo+T$, and which also occurs in the
gravitino transformation law, while in the gaugino transformation law we
have

$$Z_{A_1A_2A_3}=
  {1\over6}~T_{A_1A_2A_3}
  +6\gg_1~e^{-{4\over3}\ff}W_{A_1A_2A_3}^{(5)}
\eqno(5)$$

\noindent
with

$$W^{(5)}_{A_1A_2A_3}=
  {1\over36}\square\thinspace T_{A_1A_2A_3}
  -({1\over9}+{3\over2}h_1)~T_{B_1B_2[A_1}R^{B_1B_2}_{~~~~A_2A_3]}
  +({2\over9}+3h_1)~T_{B_1[A_1A_2}R^{B_1}_{~~A_3]}
$$$$
  +{1\over36}D_{B_1}T_{B_2[A_1A_2}T_{A_3]}^{~~B_1B_2}
  +({1\over4}+3h_1)T^{B_1B_2}_{~~~~[A_1}D_{A_2}T_{A_3]B_1B_2}
$$$$
  -({5\over54}+{5\over3}h_1){1\over{5!}}~\ee_{A_1A_2A_3B_1B_2B_3C_1C_2C_3C_4}
    T^{B_1B_2B_3}D^{C_1}T^{C_2C_3C_4}
$$$$
  -({5\over18}+{35\over6}h_1)
  {1\over{5!}}~\ee_{A_1A_2A_3B_1B_2B_3C_1C_2C_3C_4}
    T^{B_1B_2B_3}T^{DC_1C_2}T^{C_3C_4}_{~~~~D}
$$$$
  +({5\over9}+{5}h_1)~T_{B_1B_2[A_1}T^{B_2~~B_3}_{~~A_2}T_{A_3]
    B_3}^{~~~~B_1}
  +({5\over18}+{5\over2}h_1)~T_{B_1[A_1A_2}T_{A_3]B_2B_3}T^{B_1B_2B_3}
$$$$
  -({1\over108}+{7\over36}h_1)~T_{A_1A_2A_3}T^2.
\eqno(6)$$

\noindent
Obviously the field redefinitions in for instance \cite{S86}
would lead to a rather long calculation which is not needed
for the present purpose.
If we further demand that the solution fulfil

$$D_{[M}H_{NPQ]}=
  -4~\mbox{Tr}(F_{[MN}F_{PQ]})
  -\gg_1~\mbox{Tr}(R_{[MN}R_{PQ]}),
\eqno(7)$$

\noindent
with the traces defined just as the sum over the group indices,
we will also automatically satisfy the bosonic equations of motion
\cite{P91}.
In order to find a five-brane solution we split up space-time into

$$z^M\longrightarrow(y^a,x^\mu);~~~~~~~~a=0,1,..5;~~\mu=6,..9
\eqno(8)$$

\noindent
and assume a metric of the form

$$g_{MN}=
  \left(\begin{array}{lllllll}
   e^{2A}&&&&&&\\
   &-e^{2A}&&&&&\\
   &&\ddots&&&&\\
   &&&-e^{2A}&&&\\
   &&&&-e^{2B}&&\\
   &&&&&\ddots&\\
   &&&&&&-e^{2B}\end{array}
\right).\eqno{(9)}$$

\noindent
Here $A=A(r)$ and $B=B(r)$ are arbitrary scalar fields which, as well as
all the fields in the following, depend only on
$r=(\dd_{\mu\nu}x^\mu x^\nu)^{1\over2}$.
Our strategy will now be to solve (1a)-(1c) with $Z_{A_1A_2A_3}$ and
$T_{A_1A_2A_3}$ regarded as independent fields and only put the solution
into (5) afterwards.

We start by studying the dilatino equation (1b).
Just like in \cite{S90} it can be solved by defining chiral spinors
\footnote{The $\ee_{...}$'s are here defined as tensor densities;
${1\over\sqrt{g}}\ee_{...}$ are the proper tensors in respective space.}

$${1\over\sqrt{g_6}}\ee_{a_1...a_6}\gg^{a_1...a_6}\ee_\pm=
  \pm6!~\ee_\pm$$

$${1\over\sqrt{g_4}}\ee_{\mu_1...\mu_4}\gg^{\mu_1...\mu_6}\ee_\pm=
  \pm4!~\ee_\pm
\eqno(10)$$

\noindent
with $g_6=-det(g_{ab})=e^{12A}$, $g_4=det(g_{\mu\nu})=e^{8B}$,
and by putting

$$Z_{\mu_1\mu_2\mu_3}\sim
  \ee_{\mu_1\mu_2\mu_3}^{~~~~~~~\nu}\part_\nu\ff (r)~e^{C(r)},
\eqno(11)$$

\noindent
and the rest of the components to zero.
We then immediately find

$$Z_{\mu_1\mu_2\mu_3}^\pm=
  \mp{1\over3}\ee_{\mu_1\mu_2\mu_3}^{~~~~~~~\nu}
  \part_\nu \ff (r)~e^{-4B}.
\eqno(12)$$

\noindent
Note that the factor $e^{-4B}={1\over\sqrt{g_4}}$ is exactly what is
needed to make $Z_{\mu_1\mu_2\mu_3}^\pm$ a tensor in $x$-space.
Proceeding to the gravitino equation, we make a similar, but independent
ansatz:

$$T_{\mu_1\mu_2\mu_3}=
  \ee_{\mu_1\mu_2\mu_3}^{~~~~~~~\nu}\part_\nu D(r)~e^{E(r)}.
\eqno(13)$$

\noindent
The $M=a$ component of (1c) is then

$$0=
  \part_a\ee_\pm
  +{1\over2}\gg_a\gg^\mu\ee_\pm~\part_\mu A
  \mp{1\over6}\gg_a\gg^\mu\ee_\pm~\part_\mu D ~e^{E+4B}
 \eqno(14)$$

\noindent
which can be solved by making $\ee$ independent of $y^a$,
$D^\pm=\pm3A^\pm (+\mbox{constant})$, and $E=-4B$.
For $M=\mu$ we get

$$0=
  \part_\mu\ee_\pm
  \mp{1\over6}\ee_\pm~\part_\mu D
  +{1\over2}\gg_{\mu\nu}\ee_\pm ~\part^\nu(B\pm{2\over3}D).
\eqno(15)$$

\noindent
A solution is

$$\ee_\pm=
  e^{A/2}\eta_\pm
\eqno(16)$$

\noindent
with $\eta_\pm$ constant chiral and antichiral spinors,

$$B=-2A+\mbox{constant} ,
\eqno(17)$$

\noindent
and hence

$$T_{\mu_1\mu_2\mu_3}=
  \pm\ee_{\mu_1\mu_2\mu_3}^{~~~~~~~\nu}\part_\nu A~e^{8A}.
\eqno(18)$$

\noindent
The constant in (17) can be absorbed in a constant rescaling of the
coordinates and is dropped below.
The gaugino equation (1c) is now directly solved by the (anti)instanton
configuration

$$F_{\mu\nu}=
 \pm{1\over2}\ee_{\mu\nu}^{~~~\rr\ss}F_{\rr\ss} e^{8A}.
\eqno(19)$$

In the $\gg_1=0$ case we would have had
$Z_{\mu\nu\rr}\sim T_{\mu\nu\rr}$ and hence directly $A\sim\ff$.
For $\gg_1\not=0$ we have to insert our solution into (5).
After a straightforward, but cumbersome calculation, we find that only

$$W_{\mu_1\mu_2\mu_3}^{(5)}=
  \pm\ee_{\mu_1\mu_2\mu_3}^{~~~~~~~\nu}e^{8A}~[
    {1\over12}\part^\rr\part_\rr\part_\nu A
    +{1\over2}\part^\rr\part_\rr A\part_\nu A
$$$$
    -({1\over2}+9h_1)\part^\rr A\part_\rr\part_\nu A
    -({3\over2}+27h_1)\part^\rr A\part_\rr A\part_\nu A]
\eqno(20)$$

\noindent
is different from zero.
Using (20) and contracting with
$\ee^{\mu_1\mu_2\mu_3}_{~~~~~~~\mu}e^{8A}$ we get

$$\part_\mu\ff=
  -{3\over2}\part_\mu A
  -18\gg_1e^{-{4\over3}\ff}[
    {1\over12}\part^\rr\part_\rr\part_\mu A
    +{1\over2}\part^\rr\part_\rr A\part_\mu A
$$$$
    -({1\over2}+9h_1)\part^\rr A\part_\rr\part_\mu A
    -({3\over2}+27h_1)\part^\rr A\part_\rr A\part_\mu A]
\eqno(21)$$

\noindent
which can directly be integrated to

$$e^{-{4\over3}\ff+2A}=
  k
  -2\gg_1e^{2A}(
    \part^\mu\part_\mu A
    -3(1+18h_1)\part^\mu A\part_\mu A)
\eqno(22)$$

\noindent
with $k$ constant.
For $\gg_1=0$ we have to choose $k>0$, and it can be put equal to one by
once more rescaling the coordinates.
However, in the AFS case, there are also solutions for
$k\leq 0$ as we shall
see below,
 so we choose to keep $k$ as a free parameter.
Finally, our solution has to satisfy (7).
Again we find that only the $[\mu\nu\rr\ss]$
component is different from zero, and it yields

$$\part^\mu\{\part_\mu Ae^{-{4\over3}\ff+2A}
  +\gg_1e^{-4A}[
    \part^\nu\part_\nu\part_\mu A
    +2\part^\nu\part_\nu A\part_\mu A
$$$$
    -6(1+18h_1)\part^\nu A\part_\nu A\part_\mu A]\}
=
-{4\over3}e^{-4A}Tr(F_{\mu\nu}F^{\mu\nu})
\eqno(23)$$

\noindent
We insert (22), and use

$$\part^\mu=-e^{4A}\part_\mu$$
$$F^{\mu\nu}=e^{8A}F_{\mu\nu}
\eqno(24)$$

\noindent
and obtain

$$\nabla^2\bigl[
  k~e^{-6A}
  +6\gg_1~\nabla^2A\bigr]
  =-8~\mbox{Tr}~F^2
\eqno(25)$$

\noindent
Here $\nabla^2={1\over r^3}{\part\over\part r}r^3{\part\over\part r}$ is
the Laplacian in four-dimensional Euclidean space.

Comparing the expression we use to those of other authors, see for
instance \cite{CDaF,GSW} and also \cite{S86,S90},
we find agreement for $\aa '={\kk^2\over 2g^2}$ and $\gg_1=-2\aa '$ and,
in particular,

$$\mbox{Tr}(F_{\mu\nu}F^{\mu\nu})
  ={1\over{8\cdot 30}}
  \aa '~\mbox{Tr}(F_{\mu\nu}F^{\mu\nu})_{\mbox{Strominger}}
\eqno(26)$$

\noindent
Hence, our solutions are exactly the same (as they should be ) for
$\gg_1=0$ ( $\ff=-{1\over2}\ff_S$ ).
The general solution given our ansatz is then any $A(r)$ and $\ff (r)$
satisfying

$$k~e^{-6A}
  +6\gg_1~\nabla^2A
  = k~e^{-6A_0}
  +{ k'\over r^2}
  +8\aa'~{r^2+2\rr^2\over (r^2+\rr^2)^2}
\eqno(27a)$$
$$e^{{4\over3}\ff+2A}
  = k
  -2\gg_1~e^{2A}
    \bigl(\nabla^2A
          -3(1+18h_1)~(\nabla A)^2
    \bigr)
\eqno(27b)$$

\noindent
together with (9), (12), (17), and (18).
As a consistency check, as well of \cite{P91}
as of the calculations above,
the equations of motions were also studied, and found to be linear
combinations of (derivatives of) the equations (27).
\section{Analysis of the solution and discussion.}
In equations (27)
we have three, so far arbitrary, integration constants.
However, the solution is not physical, and might not even exist for all
values of $k$, $k'$, and $A_0$.
For instance, the constants have to be chosen so that $A$ and $\ff$ are real.
We will here first
restrict ourselves to a discussion of a few cases already
mentioned in the literature. All these solutions can, of course, be
extended to multi-instanton or multi-wormhole solutions in the standard
fashion.
Afterwards, we will give examples of
other exact, but mainly unphysical, solutions.
The instanton solution \cite{S90}
 (gauge solution in \cite{CHS91,C91}) has $k'=0$.
If we assume that $A$ can be written as a power series in ${1\over r}$
at infinity, we find

$$A=A_0
   +{A_1\over r^2}
   +O\biggl({1\over r^4}\biggr),
{}~~~~~~~~~r\rightarrow\infty.
\eqno(28)$$

\noindent
The $\gg_1$ part gives then no contribution at infinity and the calculation
of mass, axion charge, and the Bogomoln'yi bound still give the same result
as in the paper by Strominger.
Furthermore, since we still have the same continuous symmetries, the
zero-modes should remain unchanged, and a solution of this form
giving a real $\ff$ is then
indeed an extension of \cite{S90} to all orders in $\aa'$.

If we instead let $\rr\rightarrow0$ we obtain the generalization of the neutral
solution \cite{CHS91}, related to \cite{DL91}.
Some care must be taken in this case since it is not obvious that taking the
limit $\rr\rightarrow0$ in (27) and then solving for $A$ and $\ff$ will give
the
same result as the other way round.

Perhaps the most important example, however, is that of the symmetric
solution of \cite{CHS91,C91},
which is argued to have no higher order corrections.
In order to be able to define a Lorentz connection which equals the Yang Mills
potential we use the ``original" version of the instanton potential,
see e.g. \cite{R}.
In analogy with the authors quoted above  we thus put

$$A_\mu=
  \bar\SS_{\mu\nu}~\part_\nu ~\log\biggl(1+{\rr^2\over r^2}\biggr)
  = -6~  \bar\SS_{\mu\nu}~\part_\nu A
\eqno(29)$$

\noindent
with $\rr^2=n\aa'e^{6A_0}$.
That is

$$A=A_0
  -{1\over6}~\log\biggl(1+{\rr^2\over r^2}\biggr).
\eqno(30)$$

\noindent
Inserting this into (27a) we get

$$k~e^{-6A_0}~\biggl(1+{\rr^2\over r^2}\biggr)
  -8\aa'~{\rr^4\over r^2(r^2+\rr^2)^2}
  =
  k~e^{-6A_0}
  +{k'\over r^2}
  +8\aa'~{r^2+2\rr^2\over(r^2+\rr^2)^2},
\eqno(31)$$

\noindent
which is satisfied if we choose

$$k'=(nk-8)\aa'.
\eqno(32)$$

\noindent
Since $A$ also satisfies

$$\nabla^2A
  =6~(\nabla A)^2 ,
\eqno(33)$$

\noindent
we can eliminate the correction term in (27b) if we choose the parameter
of AFS

$$h_1={1\over18}.
\eqno(34)$$

\noindent
We obtain

$$\ff=
  -{3\over2}A
  +\mbox{constant}
\eqno(35)$$

The symmetric solution is hence a solution also to AFS for the
 choice of $h_1$ in (34). Since a particular choice of $h_1$
just corresponds to a field redefinition (4),
 this value of $h_1$ must give the same
choice of $\ff$
as in the references above.

So far, the symmetric solution is the only one we have given explicitly,
only assuming that there exist well-behaved solutions of (27) of the
neutral and gauge type too, albeit not in a closed form. The symmetric
solution is, however, not the only example of a simple solution of (27),
although the others we have found do not, in general, have an immediate
physical interpretation. Both  for the symmetric solution and for these
new ones, we have cancellations between $\mbox{tr}~F^2$
and $\mbox{tr}~R^2$ so they
do not have a proper limit as $\gamma_1\rightarrow 0$. They also have
$k\leq 0$.

In (27a) we have already implicitly assumed that $A$ has a well-defined
value, $A_0$, as $r\rightarrow\infty$, so that the metric is Minkowski
at infinity, and that $k\neq 0$. We now relax these constraints and put
$k~e^{-6A_0}=k''$. For $k''=0$ we find the solution

$$A=A_0
  +{1\over3}~\log\biggl(1+{r^2\over\rr^2}\biggr)+{1\over3}
  \log\biggl({r\over\rr}\biggr),
\eqno(36)$$

\noindent
with $k<0$ and $e^\phi<0$ if we assume (34).
Putting $k=0$ we can also add non-logarithmic terms to $A$, and we find
another solution

$$A=A_0+\frac{A_2}{r^2}
  -{1\over6}~\log\biggl(1+\frac{r^2}{\rr^2}\biggr)
  -\frac{k'}{24\alpha'} \log\biggl(\frac{r}{\rr}\biggr)-\frac{k''}
  {96\aa'}r^2,
\eqno(37)$$

\noindent
which has two free integration constants, $A_0$ and $A_2$,
and is hence the general solution for $k=0$. In order to remove the
essential singularities, we must choose $A_2$ and $k''$ as zero, and
certain values of $k'$ might then yield interesting solutions.

The effective Lagrangian of string theory should also contain higher
order terms multiplied by the transcendental coefficient $\zz (3)$
\cite{zeta3}.
It is very hard to imagine how these could occur within the framework
of AFS, although it has been suggested that they might depend on the
boundary conditions chosen when solving (3) to construct an effective
Lagrangian \cite{DaFRR88}.
Another, perhaps more likely, source is to note \cite{LP(T)}
 that AFS as used
here is a minimal supersymmetric extension of the anomaly-free
Einstein-Yang-Mills theory, and that it is possible to extend it, by
relaxing the constraints used in solving the Bianchi identities.
One can then accommodate precisely the superfield needed \cite{R4}
 for the $\zz
(4)~R^4$-terms. They might then act as counterterms, the precise value
of the coefficients being decided from cancellation of divergences,
as suggested in \cite{BR89}.

To correspond to the string $\beta$-functions,
all equations of AFS should then be augmented by $\zz (3)$ terms which,
for simple non-transcendental solutions such
as the ones given above, have to
be satisfied separately.
Since it is argued, using the  $\ss$-model approach, that the
symmetric solution is really a solution to the string
\cite{CHS91,C91}, we can
assume that the $\zz (3)$ equations are indeed satisfied separately
 in this  case.
Hence, a necessary condition for all ``normal" classical solutions to
string theory should be that they satisfy the AFS equations.
It would be interesting to study other solutions like the black
five-brane one \cite{HGS91}, and also to search for new exotic
compactifications, in this framework.
It would of course also be interesting to derive the full,
non-minimal AFS, and introduce the right $\zeta(3)$ coefficients,
but judging from the derivation of the equations of motion for AFS
\cite{P91}
this might require an unrealistic amount of computing power.
\\[2cm]
{\bf Acknowledgement} We thank Dieter L\"{u}st for useful
discussions.

\end{document}